\documentclass[reprint,amsmath,amssymb,aip]{revtex4-1}

\usepackage[utf8]{inputenc}
\usepackage[T1]{fontenc}
\usepackage{mathptmx}
\usepackage{etoolbox}
\usepackage{graphicx}
\usepackage{dcolumn}
\usepackage{bm}
\usepackage[version=4]{mhchem}
\usepackage{siunitx}
\DeclareSIUnit\angstrom{\text {Å}}
\DeclareSIUnit\atomicmassunit{u}
\DeclareSIUnit\elementarycharge{\text {\ensuremath {e}}}
\DeclareSIUnit\bohr{\text {\ensuremath {a}}_{0}}

\makeatletter
\def\@email#1#2{%
 \endgroup
 \patchcmd{\titleblock@produce}
  {\frontmatter@RRAPformat}
  {\frontmatter@RRAPformat{\produce@RRAP{*#1\href{mailto:#2}{#2}}}\frontmatter@RRAPformat}
  {}{}
}%
\makeatother
\begin{document}

\preprint{???}

\title[Derivative learning of tensorial quantities -- Predicting finite temperature infrared spectra from first principles]{Derivative learning of tensorial quantities -- Predicting finite temperature infrared spectra from first principles}

\author{Bernhard Schmiedmayer}
\email{bernhard.schmiedmayer@univie.ac.at}
\affiliation{%
University of Vienna,\\
Faculty of Physics and Center for Computational Materials Sciences,\\
Kolingasse 14-16, 1090, Vienna, Austria
}

\author{Georg Kresse}
\email{georg.kresse@univie.ac.at}
\altaffiliation[Also at ]{VASP Software GmbH, Sensengasse 8, 1090 Vienna, Austria}
\affiliation{%
University of Vienna,\\
Faculty of Physics and Center for Computational Materials Sciences,\\
Kolingasse 14-16, 1090, Vienna, Austria
}

\collaboration{SFB TACO}

\date{\today}

\begin{abstract}
    We develop a strategy that integrates machine learning and first-principles calculations to achieve technical accurate predictions of infrared spectra. Specifically, the methodology allows to predict infrared spectra for complex systems at finite temperatures. The method's effectiveness is demonstrated in challenging scenarios, such as the analysis of water and the organic-inorganic halide perovskite \ce{MAPbI3}, where our results consistently align with experimental data. A distinctive feature of the methodology is the incorporation of derivative learning, which proves indispensable for obtaining accurate polarization data in bulk materials and facilitates the training of a machine learning surrogate model of the polarization adapted to rotational and translational symmetries. We achieve polarisation prediction accuracies of about \SI{1}{\percent} by training only on the predicted Born effective charges.
\end{abstract}

\maketitle
\section{Introduction}
\label{sec:intro}
Infrared (IR) spectroscopy is an indispensable methodology for discerning local structures and the intricacies of functional groups within a material. It finds pervasive use across various scientific disciplines, firmly establishing itself as a key analytical instrument \cite{stuart2004infrared}. Notably, within the realm of catalysis research, IR spectroscopy is an important technique with the unique capability to elucidate surface structures and mechanisms at the molecular scale {\em in situ} \cite{vimont2010analysing,chabal1988surface,ryczkowski2001ir}. Beyond catalysis, the application of IR spectroscopy extends its reach into the domains of clinical and biomedical analyses \cite{barth2007infrared,ng1999infrared,luypaert2007near}. In the field of materials science, IR spectroscopy is also highly valuable, as it provides indirect insight on the bonding properties of the material \cite{theophile2012infrared,fernandez2012infrared,silva2012application}. 

Although experimental IR spectroscopy is a mature technique, it is not always a simple matter to relate an experimental IR spectrum to a specific local structural feature. Prior knowledge is almost ubiquitously required to relate experimental data to structures (fingerprints). For instance, the frequencies of certain functional groups often relate to the oxidation state of the group or the electronegativity of the adsorption site. Computer simulations, particularly first-principles (FP) methods, have played a vital role in establishing these relationships. However, such FP calculations are very often limited to zero-temperature simulations \cite{biczysko2018computational,jansen2021computational,bec2021current}, whereas in-situ observations of catalytic reactions almost always involve finite temperature. In this study, we harness recent advancements in machine learning (ML) techniques in combination with FP calculations to develop a method that yields highly accurate computational IR spectra at finite temperatures. 

In recent years, ML techniques have emerged as a potent new avenue in computational materials sciences \cite{behler2007generalized,bartok2010gaussian,bartok2018machine,bonati2018silicon,morawietz2016van}. Our study uses an on-the-fly learning method \cite{jinnouchi2019fly,jinnouchi2019phase,jinnouchi2020descriptors,jinnouchi2020fly} to generate transferable machine learned force fields (MLFFs). The first step is thus to harness MLFF to attain the required nano-second-long high-quality molecular dynamics (MD) trajectories. In the present work, we rely on a now fairly standard approach for the MLFF that uses rotational and translationally invariant descriptors and a kernel-based regression, although more refined approaches could be readily adopted \cite{drautz2019atomic,batzner20223,batatia2022mace}. Learning the polarization, though, is a more challenging task. The first approach to learn tensorial quantities dates back to $\lambda$-SOAP \cite{grisafi2018symmetry,wilkins2019accurate,sommers2020raman} that we have also decided to adapt in the present work. Equivariant message-passing networks would be equally suitable, but they are matter of fact closely related to the tensorial descriptors in $\lambda$-SOAP. 

The second key issue that we address in the present work is that the polarization is not uniquely determined in bulk materials \cite{king1993theory,vanderbilt1993electric,resta1994macroscopic,resta1998quantum}. This is in stark contrast to molecules, where the polarization can be determined by appropriate integration of the density \cite{gastegger2017machine}. Direct learning of the polarization hence requires some curation of the polarization data {\em e.g.} deriving the polarization from Wannier-centres imposing some continuity condition \cite{zhang2020efficient} or manual ``alignment" of the polarization data. 

The solution presented in this study to overcome this challenge involves employing derivative learning \cite{chmiela2017machine}. Specifically, we utilize the derivative of the polarization with respect to the ionic positions, the Born effective charge tensors. The hypothesis is that by computing this derivative information across numerous structures (along any relevant ``adiabatic" pathway), it becomes feasible to determine the anti-derivative, {\em i.e.}, the bulk polarization. This methodology offers a second crucial advantage: akin to the construction of MLFF  where learning the forces proved pivotal, the Born effective charges are significantly more expressive yet nearly as computationally efficient to calculate in solid-state codes as a single polarization.

In the next section, we summarize our methodological approach, then demonstrate the feasibility of derivative-based learning for the water dimer, and discuss results for liquid water, where we find excellent agreement with the experiment only for a functional including van der Waals corrections. Lastly, we demonstrate very good agreement for the IR spectrum of an organic perovskite with experimental data. We finish with discussions and conclusions, and mention points worthwhile an improvement.

\section{Method}
\label{sec:method}
\subsection{General remarks:}
\label{sec:energy}

In general, the energy of a molecule or material in the presence of an electric field is described
by 
\begin{equation}
    E(\mathbf{x} , \mathcal{E} )= E_{\rm KS}(\mathbf{x}) - \mathcal{E} \cdot \mathbf{P}(\mathbf{x}, \mathcal{E})
    \label{equ:energy}
\end{equation}
where $\mathcal{E}$ is the electric field, $\mathbf P$ the polarization, and $E_{\rm KS}$ is the Kohn-Sham energy at zero field for the atoms at the position $\mathbf x$ \cite{umari2002ab}.  There is an implicit dependence of the Kohn-Sham energy on the electronic field, as the orbitals need to be determined by minimizing the energy in the presence of the field, but thanks to the variational properties and the Hellman-Feynman theorem, variations of the orbitals can be neglected for first derivatives. To calculate second derivatives, only the first derivatives of the orbitals are required. Obviously, the first derivative of the energy with respect to the electric field yields the polarization $\mathbf{P}$. The second derivative of the energy with respect to the field corresponds to the electronic polarizability, and the second mixed derivative with respect to the electric field and the positions yields the Born effective charge tensor:
\begin{equation}
 \mathbf{Z}^* = \frac{\partial^2 E(\mathbf{x}, \mathcal{E})}{\partial \mathbf{x} \, \partial \mathcal{E}}.
\label{equ:mixederiv}
\end{equation}
This is a second-rank cartesian tensor. In the present work, we set out to learn the polarization as a function of the positions and neglect the dependence of the polarization on the electric field (implicitly assuming zero electric field).  If we were to evaluate the energy using Eq. \eqref{equ:energy} this would only be correct to linear order in the field. 

\subsection{Green-Kubo relation:}
\label{sec:Green-Kubo}
Using the Green-Kubo formalism, the ionic contribution to the polarizability, denoted as $\chi(\omega)$, is directly proportional to the Fourier transform of the autocorrelation function of the polarization $\mathbf{P}$ and its time derivative and can be expressed as follows (SI units)\cite{kubo1957statistical,zwanzig1965time,sangalli2012pseudopotential}:
\begin{equation}\label{eq:IIR}
    \chi_{\mu,\nu}(\omega) = \frac{\beta}{V\epsilon_0} \int_0^T \langle \mathbf{P}_{\mu}(0) \dot{\mathbf{P}}_{\nu}(t) \rangle e^{-i (\omega-i \delta) t} \mathrm{d}t.
\end{equation}
Here $\mu$ and $\nu$ represent Cartesian indices, $V$ is the volume, $\beta$ is the inverse temperature, $\epsilon_0$ the vacuum permittivity, $\omega$ the vibrational frequency, and $\delta$ denotes a complex shift causing a Lorenzian broadening. It is necessary that the product $T \delta$  is small to avoid truncation artefacts. The time derivative of the polarization $\dot{\mathbf{P}}$ can be written as
\begin{equation}
    \frac{\partial \mathbf{P}}{\partial t} = \frac{\partial \mathbf{P}}{\partial \mathbf{x}} \frac{\partial \mathbf{x}}{\partial t} = \mathbf{Z}^{*} \dot{\mathbf{x}}.
\end{equation}
To determine Eq. \eqref{eq:IIR} accurately three prerequisites exist: first, accurate velocities $\dot{\mathbf{x}}$ {\em i.e.} high-quality MD trajectories to describe the time evolution of the system, second, long simulation times $T$, and third, a reliable method to model the polarization and the Born effective charge tensors.

In infrared reflectivity or transmission experiments, an absorbance $\alpha$ is measured and specified (the absorbance is proportional to $\omega$ times the spectral function of the polarizability). We report the product of the absorbance $\alpha(\omega)$ and the refractive index $\eta(\omega)$, isotropically averaged over the three diagonal components of the polarizability tensor:
\begin{equation}
    \alpha(\omega) \eta(\omega) = \frac{\beta }{3 V c \epsilon_0} \mathfrak{Re}\int_0^T \langle \dot{\mathbf{P} }(0) \cdot \dot{\mathbf{P}}(t) \rangle e^{-i (\omega-i \delta) t} \mathrm{d}t.
\end{equation}
In this equation, $c$ is the speed of light. We note that one can also autocorrelate $\mathbf{P}(0)$ with $\mathbf{P}(t)$, or $\mathbf{P}(0)$ with $\dot{\mathbf{P}}(t)$, adding simultaneously factors $\omega^2$ and $-i \omega$, respectively. We have tested all three approaches and found identical results for the three versions.
\subsection{Molecular dynamics simulations:}
\label{sec:md}
For disordered materials, the required long simulation times are unattainable using first principles simulations, and hence surrogate models are required. To obtain FP data for constructing the MLFF, we employed the Vienna {\em Ab initio} Simulation Package (VASP) \cite{kresse1996efficiency,kresse1996efficient,kresse1999ultrasoft}. Both the training of the MLFF and its subsequent application were conducted using the ML framework integrated in the VASP code \cite{jinnouchi2019phase,jinnouchi2019fly,jinnouchi2020descriptors}. It is essential to emphasize that the quality of an MLFF is not only dependent on the underlying ML algorithm but also the quality and representativeness of the training dataset concerning the problem at hand \cite{li2013permutation,botu2015adaptive}. Ideally, the training dataset should comprehensively cover all relevant regions of the potential energy surface, and this coverage should be compact to minimize the necessity for computationally expensive FP calculations. To accomplish this, we have employed the on-the-fly learning scheme integrated within VASP.

The training dataset was curated from a collection of multiple MD trajectories. These trajectories were obtained  using Langevin thermostats \cite{allen2017computer,hoover1982high,evans1983computer} with a fiction coefficient of \SI{10}{\per\pico\second}. To comprehensively cover all phases of interest, a series of heating and cooling runs were executed as discussed in the supplementary. Notably, the temperature range considered was slightly wider than the region of interest, a choice aimed at enhancing the stability of the MLFF. 

To obtain an IR spectrum, we used multiple MD trajectories within a micro-canonical ensemble to avoid artefacts caused by thermostats. The initial configurations and velocities for these individual MD trajectories were drawn from an isothermal-isobaric ensemble. The final IR spectrum was then computed as the average of the individual IR spectra obtained from these trajectories, providing a statistically accurate representation of the system's vibrational modes. We note that IR spectra are only sensitive to the long-range vibrational modes, {\em i.e.} at zero temperature the calculations can be performed using the unit cell. However, at finite temperature, supercells are required to account for the finite temperature disorder. We checked carefully that the IR spectra are cell-size converged for all cases reported here. 

\subsection{Polarization model}
\label{sec:polarization}
As highlighted in the introduction, obtaining the polarization for bulk systems presents inherent challenges. This is due to an undetermined modulo resulting from the absence of a unique phase origin or reference point \cite{king1993theory,vanderbilt1993electric,resta1994macroscopic,resta1998quantum}. To address this problem, we use derivative learning. Specifically, the polarization $\mathbf{P}$ of a system can be understood as the antiderivative of the Born effective charge tensor $\mathbf{Z}^{*}$ \cite{wang1996polarization}. The Born effective charge tensor for ion $i$ is mathematically expressed as
\begin{equation}
    \mathbf{Z}^{*}_{i,\alpha\beta}=Z_{J(i)} \delta_{\alpha,\beta} + V\frac{\partial \mathbf{P}_\alpha^\text{elec} }{\partial  \mathbf{x}_{i\beta}}\Bigg |_{\mathcal{E}=0}.
\end{equation}
Here, $Z_{J(i)}$ represents the bare ionic charge of the $i$th ion and $J(i)$ signifies the atomic species, $V$ stands for the volume of the unit cell, and $\mathbf{P}_\alpha^\text{elec}$ corresponds to the Cartesian component $\alpha$ of the macroscopic electronic polarization. In the equation, $\mathbf{x}_{i\beta}$ denotes the position of the $i$th ion along the Cartesian component $\beta$. 

The central idea is that the surrogate ML model describes the polarization $\mathbf{P}_\alpha^\text{elec}$. However, we aim to avoid training this model on the polarization, but instead train the model on derivative data $\mathbf{Z}^{*}$. 
Clearly, the polarization must transform like a vectorial quantity under rotations. To this end, we employ ridge regression, with a linear kernel function $K$, constructed using the covariance of 3-dimensional descriptors $D^{\mu}_{n}(\mathcal{X})$. Here, $\mathcal{X}$ represents an atomic environment, $\mu$ corresponds to a Cartesian component, and $n$ is the feature dimension. The linear kernel function is defined as follows: 
\begin{equation}
    K_{\mu\nu}(\mathcal{X},\mathcal{X}') = \sum_{n} D^{\mu}_{n}(\mathcal{X})D^{\nu}_{n}(\mathcal{X}').
\end{equation}
This equation captures the similarity between atomic environments $\mathcal{X}$ and $\mathcal{X}'$. To describe the atomic environment, we utilized the $\lambda$-SOAP (Smooth Overlap of Atomic Potentials) descriptors developed by Grisafi {\em et al.} \cite{grisafi2018symmetry}. These descriptors conserve the rotational symmetry of tensorial quantities:
\begin{equation}
    \hat{S}D^{\mu}_{n}(\mathcal{X})=D^{\mu}_{n}(\hat{S}\mathcal{X}),
\end{equation}
where $\hat{S}$ is a generalized symmetry operator of the $SO(3)$ group. The descriptor is tailored to describe the surroundings of an atom. In the present work we use two- and three-body descriptors and Bessel functions as radial basis sets, as in the original MLFF implementation of VASP \cite{jinnouchi2019fly,jinnouchi2019phase}. To ensure smoothness in derivatives and avoid abrupt discontinuities, we incorporated a Behler and Parrinello cutoff function \cite{behler2007generalized}. The radial cutoffs are typically set to \SI{5.5}{\angstrom}.

The polarization $\mathbf{P}$ of a configuration, characterized by atomic environments $\mathcal{X}_j$, is determined using descriptors $D^{\nu}_{n}(\mathcal{X}_{I_\text{ref}})$ of reference atomic environments $\mathcal{X}_{I_\text{ref}}$ following the equation:
\begin{equation}
    \mathbf{P}_\alpha = \sum_{\nu I_\text{ref}}\omega_{I_\text{ref}}^\nu \sum_{jn} D^{\alpha}_{n}(\mathcal{X}_j)D^{\nu}_{n}(\mathcal{X}_{I_\text{ref}}).
\end{equation}
Here $\omega_{I_\text{ref}}^\nu$ represents weights that are paired with each reference descriptor $D^{\nu}_{n}(\mathcal{X}_{I_\text{ref}})$. These weights are determined through derivative learning of the Born effective charge tensor, as given by the equation:
\begin{equation}
    \mathbf{Z}^{*}_{\alpha\beta}(i) = \sum_{\nu I_\text{ref}}\omega_{I_\text{ref}}^\nu \sum_{jn} \frac{\partial D^{\alpha}_{n}}{\partial \mathbf{x}_{i\beta}}(\mathcal{X}_j)D^{\nu}_{n}(\mathcal{X}_{I_\text{ref}}).
\end{equation}
The weights are obtained by utilizing a linear regression model using the least squares method, allowing for simultaneous calculation of $\omega_{I_\text{ref}}^\nu$ across all training configurations. We use sparse regression, {\em i.e.} reduce the number of kernel-basis functions $\mathcal{X}_{I_\text{ref}}$ using farthest point sampling.

Our investigation revealed a substantial improvement in the quality of the fit by specially treating the diagonal elements of the Born effective charge tensor. To achieve this improvement, we applied a preprocessing step that involved subtracting the mean value of the diagonal elements of the Born effective charge tensor for each atomic species $\bar{Z}^*_J$ before the training process. Here, $J$ represents the atomic species. Subsequently, we add back to  the polarization $\mathbf{P}$ the following term:
\begin{equation}
    \bar{\mathbf{P}}_\alpha = \sum^{N_\text{atom}}_{i=1} \bar{Z}^*_{J(i)} \mathbf{x}_{i\alpha}.
\end{equation}

The computation of Born effective charges was carried out using density functional perturbation theory (DFPT) \cite{wu2005systematic,gajdovs2006linear,perez2015vibrational} by computing the static ion-clamped dielectric matrix, as discussed by Baroni and Resta \cite{baroni1986ab} and Gajdoš {\em et al.} for the projector-augmented wave (PAW) method \cite{gajdovs2006linear}. It is noteworthy that the calculation of the Born effective charge tensor is computationally three times more intensive than a comparable Density Functional Theory (DFT) groundstate calculation. Essentially, it requires calculating the orbitals' first derivative with respect to the three directions of the external fields $\mathcal{E}$, which suffices to obtain the mixed derivatives as defined in Eq. \eqref{equ:mixederiv}. Alternatively, one could also determine the first derivative of the orbitals with respect to the positions, and then predict the mixed second derivatives, however, this scales linearly with system size and is thus computationally more involved (the results for the Born effective charges are independent of the order of differentiation). To the best of our knowledge, any electronic structure code can determine the Born-effective charges via the orbital derivative with respect to the field. The present approach is therefore applicable to most first-principles codes. The inclusion of derivatives provides considerably more information, in particular, an additional $3N_{\text{atom}}$ of data, so that only a small number of FP calculations are required to achieve a high degree of accuracy as demonstrated below.

\subsection{Results and Discussion}
\label{sec:results}
To demonstrate the versatility of the developed methodology, we have applied it to three distinct systems. First, we consider the water dimer \ce{2(H2O)}. In the case of molecules, the polarization is a well-defined property. Second, we examine water. Lastly, we extend our analysis to a complex solid state system, focusing on an organic perovskite \ce{MAPbI3}, since it exhibits sizable anharmonicities.

\subsection{Water dimer}
\label{sec:waterdimer}
We start our analysis with a water dimer \ce{2(H2O)}. Given that molecules possess a well-defined polarization, we can make a comparative assessment between the derivative learning approach and the conventional method of directly learning the polarization. This provides a valuable benchmark for assessing the reliability of the methodology.

The training and test configurations were extracted from an MD trajectory. We commenced with a water dimer placed in a simulation box with ample vacuum space. One of the oxygen atoms was constrained using selective dynamics to anchor the system. To simulate thermalization, we executed a heating run, spanning temperatures from \SIrange{10}{320}{\kelvin}, utilizing a Langevin thermostat. Importantly, the MD run was enhanced with on-the-fly machine learning to speed up the computational process. For the FP calculations, we opted for a revised Perdew-Burke-Ernzerhof (RPBE) functional \cite{hammer1999improved} with van der Waals (vdW) dispersion energy corrections of Grimme {\em et al.} with zero-damping function \cite{grimme2010consistent} (RPBE-D3). This choice ensured that our calculations account for vdW interactions.

During the MD, the mass of the hydrogen atom was increased to \SI{8}{\atomicmassunit}, to allow for larger time steps of \SI{1.5}{\femto\second}. Out of a total of \num{200000} MD time steps, we selected \num{1000} uniformly distributed configurations for the computation of the polarization.  For calculating the Born effective charge tensor \num{150} configurations were chosen. To determine polarization, we integrated the total (electronic and ionic) charge times the position operator by switching on dipole corrections in VASP (see {\em e.g.} Refs. \cite{makov1995periodic,neugebauer1992adsorbate}). To improve the accuracy of the Born effective charge tensor, we used a strict convergence criterion of \SI{1e-7}{\electronvolt} for the electronic self-consistency loop and large cells. To confirm the internal consistency between polarization and Born effective charges, we also calculated numerical derivatives of the polarization. These derivatives showed an excellent agreement with the Born charges with a root mean square error (RMSE) less than \SI{5e-6}{|\elementarycharge|}, confirming the reliability of the resulting database.

\begin{figure}[htb]
\includegraphics{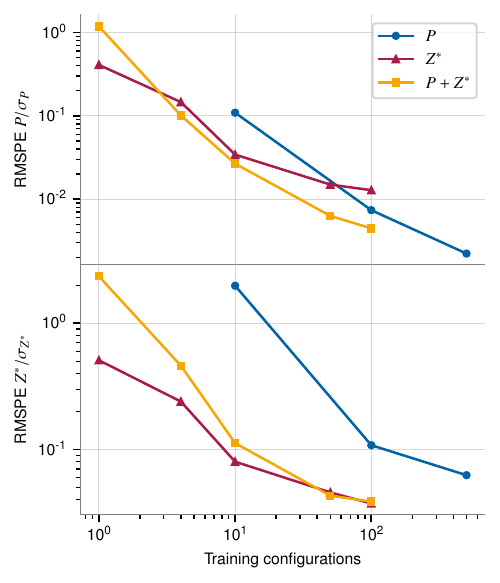}
\caption{\label{fig:learning}Learning curves for the polarization ($P$) and Born effective charges ($Z^*$) of a water dimer. Errors on the test set are expressed in root mean square percentage errors (RMSPE), where the RMSE is normalized by the standard deviation ($\sigma$) of the test set to yield a dimensionless quantity (note that $10^{-1}$ and $10^{-2}$ at the y-axis correspond to \SI{10}{\percent} and \SI{1}{\percent}, respectively). The blue line (circle) denotes training solely on polarization ($P$), and the red line (triangle) represents training solely on Born effective charges ($Z^*$). The yellow line (square) demonstrates combined learning of polarization and Born effective charges ($P+Z^*$). }
\end{figure}

To construct the learning curves, we used a dataset consisting of \num{1000} polarization calculations. The dataset was split in half by alternately selecting configurations for training and validation. Additionally, we selected \num{50} Born effective charge calculations as validation data, chosen to be uniformly distributed across the dataset of \num{150} Born effective charge tensor calculations. The remaining \num{100} Born effective charge calculations served as the training dataset. During the combined training process, the polarization data was weighted ten times higher than the Born effective charge data. To determine the optimal number of fitting parameters, we selected the number of kernel functions that minimized the error in polarization prediction. It is worth noting that, as proposed by Cortes {\em et al.} for regression \cite{cortes1993learning}, there should be a relationship between the RMSE and the number of training data, characterized by a power-law decay. The learning curves, which provide insights into the model's performance, are presented in Fig. \ref{fig:learning}.

The learning curves demonstrate the power-law relation between the RMSE and the number of training configurations. However, it is important to note that with a large number of training data points, the improvement in the mean squared errors seems to plateau somewhat, indicating that our linear regression with two- and three-body descriptors will likely yield some residual (but acceptable) model errors. Overall, the most favourable results are achieved through combined learning, with percentage errors below \SI{0.5}{\percent} and \SI{3}{\percent} for the polarization and the Born effective charges, respectively. Crucially, learning only Born charges from \num{100} training configurations also attains a high relative accuracy of approximately \SI{1}{\percent} for the polarization, and is as accurate as combined learning for the Born effective charges. We also note that there is no noticeable offset in the ML polarization compared to the FP calculations. Likely this is so since the molecule is free to rotate (and does rotate during the MD) and the surrogate model reliably determines the offset. On the other hand, attempting to train on the polarization data only requires more training data (blue line), but even with \num{500} training data points, the Born effective charges still show twice the errors as combined training does using \num{100} training data points. 

\subsection{Liquid water}
\label{sec:liquidwater}
For our second proof-of-concept, we selected liquid \ce{H2O} at room temperature. While the IR spectrum of water has been extensively studied and is well understood \cite{falk1966infrared,draegert1966far,walrafen1972water,hasted1985far,bertie1996infrared,schmidt2007structural}, theoretical interpretations of these spectra remain challenging \cite{madden1986infrared,bansil1986molecular,guillot1991molecular,corongiu1992molecular,silvestrelli1997ab,auer2008ir,xu2019first,medders2015infrared}. Even with the use of modern DFT methods, accurately reproducing the experimental properties of water is a complex task and often yields results that are far from accurate \cite{gillan2016perspective}. As a result, this system serves as an ideal test case for validating the reliability and effectiveness of the methodology as well as gleaning some insight on the underlying dynamics of water.

We chose the training configurations from an MD trajectory using an on-the-fly learning scheme. The simulation was conducted within a cubic box with periodic boundary conditions, containing a total of \num{64} \ce{H2O} molecules. The lattice constant of the cubic box was adjusted to attain a density of approximately \SI{997}{\kilogram\per\meter\cubed}, closely resembling the density of water at room temperature \cite{MTanaka_2001}. Data was collected during multiple heating and cooling runs conducted within a canonical MD ensemble. We employed various temperatures, spanning from \SIrange{270}{420}{\kelvin}. The MLFF was trained using FP data obtained from DFT calculations. Specifically, we again employed the RPBE-D3 functional but found it necessary to use hard PAW potentials and a cutoff of \SI{800}{\electronvolt} to obtain accurate stretch frequencies \cite{de2023comparing}. Additionally, we trained an MLFF using SCAN \cite{sun2015strongly} to determine whether SCAN offers a reasonable description of the water dynamics.

\begin{figure}[htb]
    \centering
    \includegraphics{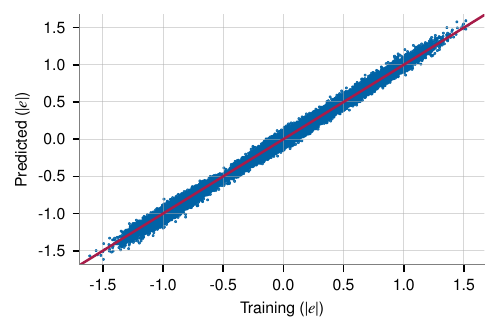}
    \caption{\label{fig:H2O_Scatter} Scatter plot of the trained and predicted Born effective charges for water.}
\end{figure}
The training configurations for the tensorial machine learning framework were once more chosen from a canonical MD ensemble. This ensemble was maintained at a constant temperature of \SI{298.2}{\kelvin} and controlled by a Langevin thermostat. A total of \num{10000} MD steps, accelerated by the MLFF, were executed. From the MD trajectory, \num{100} configurations were uniformly selected as the training dataset for the Born effective charges. A scatter plot of the trained and predicted Born effective charges is shown in Fig. \ref{fig:H2O_Scatter}.

\begin{figure}[htb]
    \centering
    \includegraphics{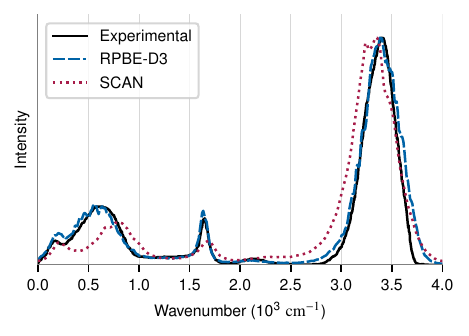}
    \caption{\label{fig:H2O_IR} Experimental and computational IR spectra for liquid water. The experimental reference data is from Ref. \cite{bertie1996infrared}}.
\end{figure}
The computational IR spectrum presented alongside experimental data in Fig. \ref{fig:H2O_IR} was computed by averaging the results over \num{20} individual IR spectra. Each of these individual spectra was calculated from a micro-canonical MD trajectory. For each MD run, we initiated the simulation from an uncorrelated starting configuration, ensuring the appropriate average corresponding to room temperature. In each run, we performed a total of \num{100000} MD steps, with a time step of \SI{0.25}{\femto\second}. This strategy of conducting multiple calculations from uncorrelated starting configurations was employed to enhance the statistical accuracy of our results. Instead of Lorenzian broadening, a Gaussian filter was applied before performing the Fourier transform. This improves the feature sharpness in the computed IR spectra.

As illustrated in Fig. \ref{fig:H2O_IR}, the present methodology allows the computation of the IR spectrum of liquid water with a remarkable level of agreement with experimental data. There are two important conclusions to draw from this result. That the intensity follows so closely the experimental data, is related to an accurate description of the Born effective charges. Since we train on the Born effective charges, the good agreement is likely not astonishing. Second, RPBE+D3 yields an excellent description of the dynamics of liquid water, both for the high-frequency stretch, but also for the medium-frequency bending motions. Most important are the lower frequency modes that are related to intra-molecular motions. It is important to note that the results for the SCAN functional are significantly worse for these modes. Specifically, the low-frequency mode around \SI{500}{\per\centi\meter} is wrong by almost \SI{40}{\percent}  indicating serious deficiencies in the description of the molecular motion of water molecules en-caged by the four surrounding water molecules.

Previous works, such as those by Sommers {\em et al.} \cite{sommers2020raman}, Zhang {\em et al.} \cite{zhang2020efficient}, and Gastegger {\em et al.} \cite{gastegger2017machine}, have undertaken similar approaches utilizing MD trajectories in conjunction with symmetry-preserving machine learning frameworks to obtain Raman spectra for water and IR spectra for molecules. In these prior studies, the focus has been on learning the polarization directly. This required significantly more training data and a careful calculation of the polarization in order to avoid any discontinuities. 

\subsection{Anharmonic solids}
\label{sec:solid}
Our final test system is the organic-inorganic halide perovskite, methylammonium lead iodide (\ce{MAPbI3}). This material has been the subject of numerous experimental and theoretical studies, including state-of-the-art vibrational studies  \cite{weber1978ch3nh3pbx3,onoda1990calorimetric,MTanaka_2001,kawamura2002structural,stoumpos2013semiconducting,baikie2013synthesis,bokdam2016role,schuck2018infrared,jinnouchi2019phase}. Notably, \ce{MAPbI3} undergoes three entropy-driven phase transitions: from an orthorhombic phase to a tetragonal phase at \SI{160}{\kelvin}, and from the tetragonal phase to a cubic phase at \SI{330}{\kelvin}. The thermodynamic nature of this material makes it challenging to model the IR spectra of the tetragonal phase using traditional \SI{0}{\kelvin} methods like DFPT \cite{wu2005systematic,gajdovs2006linear,perez2015vibrational}. Therefore, \ce{MAPbI3} serves as a valuable test case for validating the reliability of the scheme for strongly anharmonic solids.

\begin{figure*}[htb]
    \centering
    \includegraphics{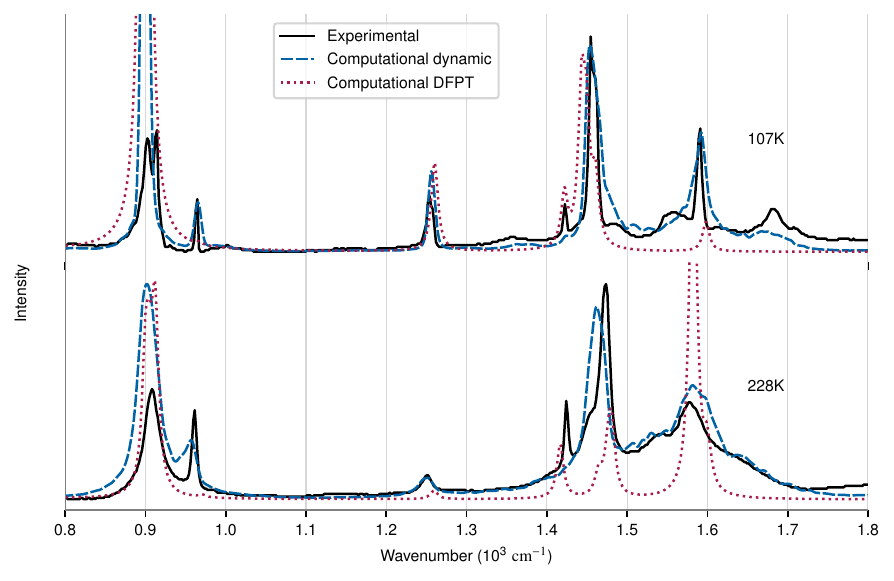}
    \caption{\label{fig:MAPI_IR} Experimental and computational IR spectra for the orthorhombic and tetragonal phases of \ce{MAPbI3}. The vibrational frequencies of the computational IR spectra have been red-shifted by \SI{1.5}{\percent} for alignment with experimental data. The experimental reference data is from Ref. \cite{schuck2018infrared}.}
\end{figure*}
The training process for the MLFF applied to \ce{MAPbI3} closely mirrored the methodology employed for water and previous studies of MAPbI3 \cite{jinnouchi2019phase}. Multiple heating runs, encompassing the two phases--- orthorhombic, and tetragonal ---were conducted using Langevin thermostat-driven MD simulations. The temperature range spanned from \SIrange{80}{430}{\kelvin}. Initially, fixed cell volumes were used, followed by simulating an isothermal-isobaric ensemble using the Parrinello-Rahman method \cite{parrinello1980crystal,parrinello1981polymorphic}. The training was performed using a strongly-constrained and appropriately normed (SCAN) meta-gradient corrected functional \cite{sun2015strongly}. We note that we found in previous studies that SCAN is better suited for the simulation of MAPbI3 \cite{bokdam2017assessing} than say RPBE-D3, as RPBE-D3 does not account for screening of the vdW interactions by the strongly polarizable cage atoms.

In Fig. \ref{fig:MAPI_IR}, we present the IR spectra for the orthorhombic phase at \SI{107}{\kelvin} and the tetragonal phase at \SI{228}{\kelvin}, alongside experimental results for comparison. The spectrum of the well-ordered orthorhombic phase and the tetragonal phase was calculated using a $4\times4\times4$ supercell for better statistics and to allow for some reordering of the MA molecules. The starting configurations for the individual MD runs were chosen from an isothermal-isobaric ensemble. Aside from using starting configurations with varying cell vectors, the procedure for calculating the IR spectrum closely mirrors the one described for water.

The analysis of the computed IR spectra reveals excellent agreement with the experimental data but also some small discrepancies, particularly, in the intensities of the individual peaks. Notably, the peaks around \SI{900}{\per\centi\meter} appear with higher intensity.  The experimental reference spectra, obtained from a single crystal \cite{schuck2018infrared}, may also be influenced by surface effects and crystal structure orientation, contributing to variations in intensity between computational and experimental results.

We start with a comparison for the orthorhombic low-temperature phase (\SI{107}{\kelvin}). The agreement between the DFPT and the finite temperature simulation is very good, but there are some marked improvements in the finite temperature data. The first peak around \SI{900}{\per\centi\meter} shows a double peak in both the DFPT and FT simulation, in agreement with the experiment. The peak at \SI{970}{\per\centi\meter} is completely missing in the DFPT data but visible in the finite temperature data. It is a result of anharmonic interactions. The shoulder at \SI{1420}{\per\centi\meter} is pronounced using DFPT but washed out at finite temperature. This peak corresponds to the \ce{CH3}-bending motions \cite{perez2015vibrational}. In the supplementary, we show that using the ionic charges only, this peak is visible in the spectrum, but the intensity of the peak is overestimated. Calculating the electronic contribution (el) only shows an almost identical peak, however, the phase of the electronic polarization is opposite to the ionic contribution, so when autocorrelating the sum of the electronic and ionic dipoles, the peak is strongly suppressed. This brings better overall agreement with the experiment, where the peak is also weak, but likely our Born effective charges are not quite sufficiently accurate (linear regression). The physics behind the vanishing of the peak is fairly simple: in this particular mode the H atoms move orthogonal to their bond direction, a direction in which the total Born effective charges $\mathbf{Z}^*$ are small to start with (electronic contribution cancels ionic one). Furthermore, the movement of three hydrogen atoms is concerted being close to a helicopter motion which reduces the IR intensity further.

In the experiment, we see quite some intensity between \SI{1450}{\per\centi\meter} and \SI{1600}{\per\centi\meter}, which is also nicely reproduced by the FT simulations. We note that this frequency range becomes even more populated in the tetragonal phase in the FT simulations, and this population is a result of the strongly anharmonic rattling motion of the molecules in the cage in turn affecting the bending motion of the hydrogens.

The tetragonal phase spectrum (\SI{228}{\kelvin}) is also in excellent agreement with the experimental spectrum. We note that the DFPT spectrum now shows many deficiencies, with a complete lack of peaks at \SI{950}{\per\centi\meter} and a tiny peak around \SI{1250}{\per\centi\meter}, as well as sharp features around \SI{1580}{\per\centi\meter}. The FT data resolves these issues, albeit the two main peaks around \SI{900}{\per\centi\meter} and \SI{1480}{\per\centi\meter} are somewhat too broad and washed out, and again the peak around \SI{1480}{\per\centi\meter}  is suppressed by the electronic contribution.

\section{Conclusion}
\label{sec:conclusion}
The present study showcases the effectiveness of a computational framework combining first-principles simulations with machine-learning methodologies. Machine-learned force fields can be used to access timescales that were previously very difficult to attain. This allows us to obtain vibrational spectra with excellent statistical accuracy, which would be very expensive to calculate without force fields. The main advance of the present work is, however, that we learn the polarization from its derivative the Born effective charges. As mentioned in the main text,  in VASP--- but likely so in any plane wave-based code ---the calculation of the Born-effective charges via the first derivative of the orbitals with respect to external fields is only roughly three times more costly than a groundstate Kohn-Sham calculation. Learning of force fields using first-principle nuclear derivatives is now ubiquitous, and learning the polarization from its nuclear derivative is a natural extension to this idea. Crucially, we have shown that the inclusion of polarization data, which is difficult to determine without some arbitrary modulus, is not required. As the polarization is the anti-derivative of the Born-effective charges, it can be directly calculated from the machine-learning model as long as derivative data is supplied along all relevant adiabatic pathways. We successfully applied this approach to challenging scenarios, including water and the organic-inorganic halide perovskite \ce{MAPbI3}. In both cases, our method demonstrates excellent agreement with experimental results, highlighting its capacity to capture the vibrational properties of diverse materials.

We finish with a few comments on further developments. In the present work, we have only used linear regression with two-body and three-body descriptors. Although the prediction accuracies are good (condensed matter) to excellent (molecules), we feel that the inclusion of higher body-order terms, or non-linear kernel-based regression might improve the predicted Born-effective charges further. Furthermore, the subtraction of a diagonal component from the Born-effective charges, albeit not particularly cumbersome, seems somewhat unsatisfactory, and one would prefer to avoid it. 

Overall, the present methodology is already very robust and can be readily applied to many relevant problems.  Further research could be directed towards infrared simulations of more complex adsorbates on surfaces or of water interacting with electrodes.

\section{Acknowledgements}
This research was funded in whole by the Austrian Science Fund (FWF) 10.55776/F81. For open access purposes, the author has applied a CC BY public copyright license to any author accepted manuscript version arising from this submission. The computational results presented have been achieved in part using the Vienna Scientific Cluster (VSC).

\bibliography{ref.bib}
\end{document}